\documentclass[12pt]{article}
\usepackage[latin1]{inputenc}
\usepackage{amssymb}
\usepackage{subfigure}
\usepackage{graphicx}
\usepackage{float}

\usepackage{amsmath}
\usepackage{color}

\newcommand{\bb}{\begin{equation}}
\newcommand{\ee}{\end{equation}}
\newcommand{\bega}{\begin{eqnarray}}
\newcommand{\ega}{\end{eqnarray}}
\newcommand{\begae}{\begin{eqnarray*}}
\newcommand{\egae}{\end{eqnarray*}}

\newcommand{\h}{\hspace*{4ex}}

\newcommand{\cent}{\centerline}
\newcommand{\vs}{\vspace*}

\begin{document}

\baselineskip 0.5cm

\begin{center}

{\large {\bf Propagation of Airy beams with ballistic trajectory passing through the Fourier transformation system } }

\end{center}

\vs{0.2 cm}

\cent{Rafael A. B. Suarez$^{\: 1}$, and Marcos R. R. Gesualdi$^{\: 1}$}

\vs{0.2 cm}

\centerline{{\em $^{\: 1}$ Universidade Federal do ABC, Av. dos Estados 5001, CEP 09210-580, Santo Andr\'e, SP, Brazil.}}

\vs{0.5 cm}

{\bf Abstract  \ --} \ In this work, we present the theoretical and experimental study of the propagation of Airy beams when passing through two convex lenses with different focal distances. The theoretical analysis is presented from a matrix optical model for an Airy beam. The experimental results are obtained in a holographic setup for experimental generation of Airy beams based on the holographic reconstruction system of computer generated holograms (CGH) implemented electronically in a spatial light modulator (SLM) and a 4f Fourier transformation system. These are in agreement with the theoretical prediction. The novelty of this work went to show the evolution of the Airy beam propagation for different convex lenses in a 4f system and its dependence on focal distances. These are important for the control and manipulation of Airy beams; and, they show new possibilities of applications of the Airy beams in optical tweezers, optical communications and optical guidance. \\


\vs{0.5 cm}

\h {\em\bf 1. Introduction} --- Airy beams (AiBs) are special beams that presents interesting features such as the ability to remain diffraction-free over long distances while they tend to freely accelerate during propagation, this was showed theoretically and experimentally by Siviloglou $et$ $al$ \cite{Sivilo2007,Siviloglou2007} and has attracted great interest in optics and atomic physics. The origin of these particular features, explained by Berry and Balazs in 1979, is due to a non-trivial solution of the Schrodinger equation in quantum mechanics for a free particle and the caustic envelope overlapped by the superposition of a plane wave \cite{Berry1979}. These properties of the Airy beams have also inspired prominent research interests and potential applications in  optical tweezers \cite{Zhang2011,Cheng2014,Zheng2011,Cao2011}, optical communications \cite{Vettenburg2014}; plasma physics \cite{Polynkin2009,Klein2012}; optical microscopy \cite{Vettenburg2014}; optical Airy-Vortex beams \cite{Dai2010,Jiang2012}, and others applications in optics and photonics.

On other hand, the holographic techniques enables the information of the amplitude and phase of an object or optical wave to be recorded on a holographic recording medium \cite{hariharan1996optical}. Recently, new possibilities emerged with the development of computers and electronic devices that are ever faster and of higher resolution, such as CCD cameras and the spatial light modulators (SLMs); new laser sources; optical systems, opto-mechanical devices of excellent quality and novel photosensitive materials. Then, these factors allowed the experimental implementation of holographic systems of numerical and optical reconstruction of wavefront of objects and optical beams \cite{Tricoles1987,Vasara1989,Vieira2014,Suarez2016,Yepes2019}.

In this way it has been possible to generate and control the propagation of the special optical beams (particularly, the Airy beam),  both computationally through the variation of the parameters of the beams and computer-generated holograms (CGHs); as well as, optically from the implementation of optical systems that change the characteristics of AiB propagation without altering its particularities such as being diffraction-free and auto-acceleration \cite{Suarez2016,Morris2007}. Some works were done to see the changes occurred in the propagation of the beams of Airy when passing through lenses \cite{bandres2007airy,han2012fractional,zhou2012fractional}. However, few with the theoretical formalism and the experimental analysis of this effect using lenses of different focal distances in a holographic system of optical reconstruction of the CGHs of 1D and 2D Airy beams. This study of beam parameter resizing is very important in applications of this beam type, for example in optical tweezers. Since we still have a limitation on the resolution of SLMs  to generate beams in very small dimensions, the use of lenses becomes necessary in this type of application.

In this work the theoretical analysis of the propagation of Airy beams when passing through two convex lenses with different focal distances is presented from a matrix optical model for an Airy beam (AiB), and extended to array of airy beam (AAiBs). The experimental results are obtained in a optical setup for optical generation of Airy beams based on a holographic reconstruction system of computer generated holograms (CGHs) implemented electronically in a spatial light modulator (SLM) and an optical 4f Fourier transform system. Thus it was possible to predict and verify important parameters of the propagation of an Airy beam, such as: the transverse intensity pattern, the propagation distance maintaining its non-diffracting properties and the deflection of the Airy beam.

\h {\em\bf 2. Theoretical background} ---

The solution for Airy beams propagating with finite energy can be obtained by solving the normalized paraxial equation of diffraction in $1$D \cite{Sivilo2007,Morris2007} 
\begin{equation}
i\frac{\partial}{\partial \xi}\psi\left(s,\xi \right)+\frac{1}{2}\frac{\partial^{2}}{\partial s^{2}}\psi\left(s,\xi \right)=0\,,
\label{paraxial_equation}
\end{equation}
where $\psi$ is the scalar complex amplitude, $s=x/x_{0}$ and  $\xi=z/kx_{0}^{2}$ are the dimensionless transverse and longitudinal coordinates, $x_{0}$ its characteristic length and $k=2\pi n/\lambda_{0}$ is the wave-number of an optical wave. The equation \ref{paraxial_equation} admits a solution at $\xi=0$, given by \cite{Siviloglou2008}
\begin{equation}
\psi\left(s,0 \right)=Ai\left(s \right)\text{exp}\left(as \right)\text{exp}\left(i\nu s \right)\,,
\label{Airy_finity}
\end{equation} 
where $\text{Ai}$ is the Airy function, $a$ is a positive quantity which ensures the convergence of Eq.\ref{Airy_finity}, thus limiting the infinity energy of the Airy beams and $\nu$ is associated with the initial launch angle of this beam. The scalar field $\psi\left(s,\xi \right)$ is obtained from the Huygens-Fresnel integral, which is highly equivalent to Eq.\ref{paraxial_equation} and determines the field at a distance $\xi$ as a function of the field at $\xi=0$ \cite{Morris2007,Siviloglou2008,Goodman_2004}, that is 
\begin{equation}
\begin{split}
\psi\left( s, \xi \right)&=Ai\left(s-\dfrac{\xi^{2}}{4}-\nu \xi + ia\xi\right)\text{exp}\left[ a \left(s - \dfrac{\xi^2}{2} - \nu \xi \right)\right] \\
	& \times  \text{exp}\left[ i \left( - \dfrac{\xi^3}{12} + \left(a^2 -\nu^2 + s \right)\dfrac{\xi}{2}  + \nu s -\nu \dfrac{\xi^2}{2} \right)\right].
\end{split}
\label{Airy_1D}
\end{equation}

This equation shows that the intensity profile decays exponentially as a result of modulating it with a spatial exponential function on the initial plane $\xi=0$. The term $s_{0}=s-\left( \xi^{2}/4\right)-\nu \xi$, where $s_{0}$ denotes the initial position of the peak at $\xi=0$, defines the transverse acceleration of the peak intensity of Airy beams. \\

\begin{figure}[H]
 \centering
 \includegraphics[scale=0.30]{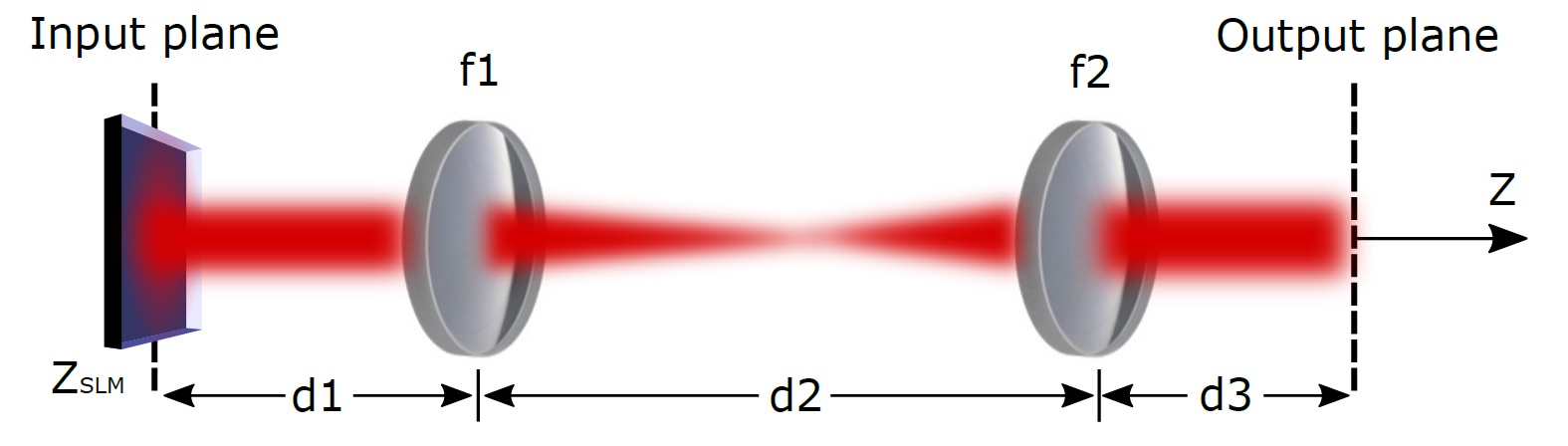}
 \caption{Lens configuration in a $4F$ Fourier filter system.}
 \label{4f_system}
\end{figure}

\textbf{\textit{Propagation of Airy beams passing through the Fourier transformation system}}. Optical system for performing the Fourier transformations is shown inf Fig.\ref{4f_system}. A 4f image system, composed by two convex lenses of focal distances $f_{1}$ and $f_{2}$, which consists of a sequence of two Fourier transforms are placed directly after the input plane. The Airy beams incidents in the input plane, and we observe the intensity distribution of the Airy beams in the output plane.

The 4f optical system is described by the following transfer matrix
\begin{equation}
M=\begin{pmatrix}
A & B\\ 
C & D
\end{pmatrix} = \begin{pmatrix}
\dfrac{-f_{2}}{f_{1}} & \dfrac{-f_{1}}{f_{2}}z\\ 
0 & \dfrac{-f_{1}}{f_{2}}
\end{pmatrix} \,,
\label{ABCD-matrix}
\end{equation} 
where we have considered $d_{1}=f_{1}$, $d_{2}=f_{1}+f_{2}$, $d_{3}=f_{2}+z$. 

The propagation of a Airy beam one-dimensional through a paraxial optical system ABCD can be described by the generalized Huygens-Fresnel diffraction integral which is given by \cite{Morris2007},
\begin{equation}
\psi\left(x,z \right)=\dfrac{1}{\sqrt{i\lambda B}}\int_{-\infty}^{\infty} \psi(\eta,z=0)\text{exp}\left[\dfrac{ik}{2B}\left(Dx^{2}-2\eta x +A\eta^{2} \right) \right] d\eta\,,
\label{generalized_Huygens_Fresnel_diffraction_integral}
\end{equation}
where $\psi(x,z=0)$ and $\psi(\eta,z=0)$ are the field in the initial and final transverse plane, respectively. The coefficients A, B and D are elements of transfer matrix corresponding to the ABCD optical system.

Substituting the Eq.\ref{Airy_finity} in \ref{generalized_Huygens_Fresnel_diffraction_integral}, we can to obtain an analytical expression for Airy beams with initial launch angle at the out put plane:
\begin{equation}
\begin{split}
\psi\left(x,z \right)&=\dfrac{1}{\sqrt{A}}\text{exp}\left(\dfrac{ikC}{2}x^2 \right)\text{exp}\left[a\left(\dfrac{x}{Ax_{0}}-\dfrac{B^2}{2A^2k^2x_{0}^4}-\dfrac{\nu B}{Akx_{0}^2}\right)\right]\\
&\times \text{exp}\left\lbrace i\left[- \dfrac{B^3}{12A^3 k^3 x_{0}^6} + \left(a^2 - \nu^2 +\dfrac{x}{Ax_{0}} \right)\dfrac{B}{2Akx_{0}^2} + \dfrac{\nu x }{Ax_{0}}-\dfrac{\nu B^2}{2A^2k^2x_{0}^4}\right] \right\rbrace \\
&\times Ai \left(\dfrac{x}{Ax_{0}}-\dfrac{B^{2}}{4A^2k^{2}x_{0}^4}-\dfrac{\nu B}{Akx_{0}^2}+\dfrac{ia B}{Akx_{0}^2} \right).
\end{split}
\label{Airy_ABCD}
\end{equation}  
From the Eq.\ref{Airy_ABCD} and \ref{ABCD-matrix} we can see that the ballistic trajectory is described by the following parabolic curve
\begin{equation}
x=-\dfrac{f_{1}^3}{f_{2}^3}\left(\dfrac{z^2}{4k^2 x_0^3} \right) - \dfrac{f_{1}}{f_{2}}\left(\dfrac{\nu z}{kx_{0}} \right). 
\label{Balisctic_Trajectory}
\end{equation} 

This results revels that the Airy beams conserve your properties when it passed through an optical system but the characteristic parameters as dimensions of the transverse intensity pattern, the characteristic length and curvature (deflection) of the Airy beam can be modified. We will see this validated and illustrated in the following sections. \\

\h {\em 3. Experimental generation of Airy beams via holographic method} --- The holographic computational methods are now a well established technique for generation and characterization of special optical beams and structured light, particularly non-diffracting beams and with orbital angular momentum \cite{Suarez2016,Yang2016,Fang2018,Yepes2019}. The computer-generated holograms of these special beams are calculated and implemented in spatial light modulators and reconstructed optically in a holographic setup. Particularly, these methods have generated experimental results of high quality and fidelity optical beams reconstruction compared to the theoretical predicted, because the holographic technique is an extremely accurate tool in the reconstruction of amplitude and phase of optical waves. Recent works using these tools has explored applications in optical tweezers for manipulation of many particles \cite{curtis2002dynamic}, in image analysis of micro-structures and biological microscopy \cite{yepes2017dynamic,Brito2013}.

Particularly, in this work from the field that described the Airy beams Eqs. \ref{Airy_1D}, we build a Computer Generated Hologram (CGH) which is optically reconstructed using a an Spatial Light Modulator (SLM).  The computer-generated hologram is calculated using an amplitude function which consists in varying the coefficient of transmission or refection of the medium  from the following amplitude function
\begin{equation}
H\left( x,y\right)= \frac{1}{2}\left\lbrace \beta\left( x,y\right) +a\left( x,y\right)cos\left[\phi\left( x,y\right)-2\pi \left(\xi x + \eta y \right)  \right] \right\rbrace \,,
\label{transmission}
\end{equation}
where $a\left( x,y\right)$ is the amplitude and $\phi\left( x,y\right)$ is phase of the complex field, $\left(\xi,\eta \right)$ is a spacial frequency of the plane wave using as reference and $\beta\left( x,y\right)=\left[1+a^{2}\left(x,y \right)\right]/2 $ is the function bias taken as a soft envelope of the amplitude $a\left( x,y\right)$. The plane wave of reference is off-axis and introduces frequencies that separate the different orders of the encoded field, see references \cite{Arrizon2005,Vieira2015}

\begin{figure}[H]
 \centering
 \includegraphics[scale=0.24]{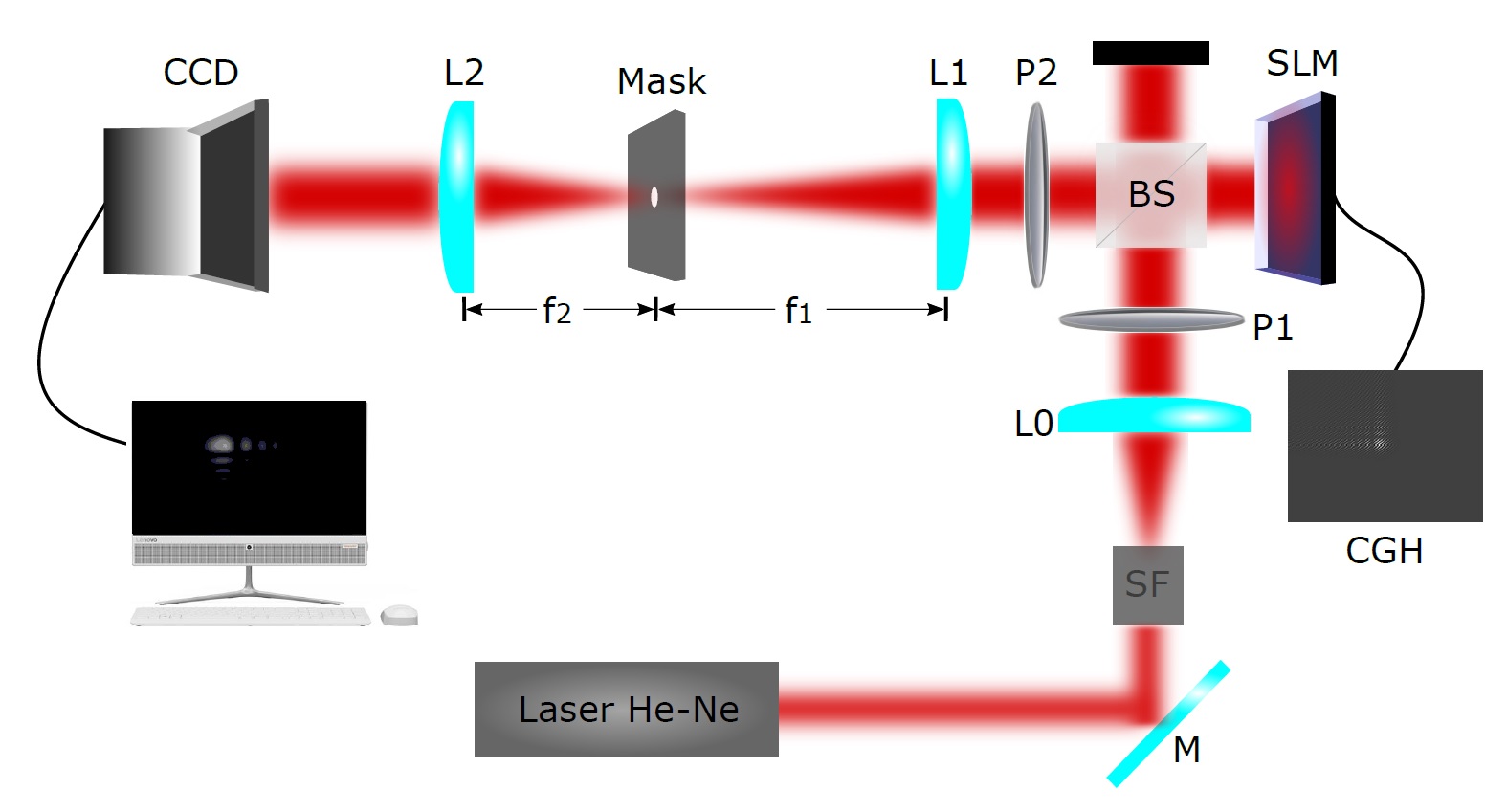}
 \caption{Experimental setup for Airy beams generation. SF is a spatial filter. Len L0, with focal lengths $75mm$ act to expand the laser beam. $\text{P1}$ and $\text{P2}$ are polarizers, which are aligned forming a angle of $0^{0}$ and $90^0$ with relation to axis $y$ of the spatial light modulator SLM. BS is a beam-splitter. Lens L1 and L2 forming a 4f system where $f_{1}=150mm$ and $f_{2}$ will take different values. ID is the mask and CCD is the camera for image acquisition.}
 \label{Setup}
\end{figure}

The experimental holographic setup for Airy beams generation is shown in the Figure \ref{Setup}.  It basically consists of a He-Ne laser beam $\left( \lambda=632.8nm\right) $, that is expanded by the spatial filters $SF$ and collimated by the lens $L0$, incident perpendicularly on the surface of the SLM ($LC-R1080$ model, of the Holoeye Photonics, with a liquid crystal display of  $1980\times 1200$ and pixel size = $8.1\mu m$) which is placed at the input plane (focus of lens $L1$). Here we use the amplitude modulation with the polarizer $P1$ ($0^{0}$) and polarizer $P2$ ($90^{0}$). A 4$f$ system is used as spatial filtering system for the Airy beams experimental generation where the lens $L1$ is fixed and we will change the lens $L2$ to study the propagation properties of the Airy passing through this system. Finally, the transverse intensity pattern of the reconstructed Airy beam is recorded by a CCD sensor "step by step" along the propagation axis.

\h {\em 4. Results and Discussion} --- For optical generation of Airy beams was used the experimental setup Fig.~\ref{Setup}. Initially, we build a CGH of the field described by Eq.~\ref{Airy_1D} from the Eq.~\ref{transmission} adopting the carrier of frequencies $\eta=\xi=\Delta p/5$ for the plane wave of reference, where $\Delta p=1/\delta p$ is the bandwidth and $\delta p$ is the individual pixel size ($8.1\mu m$). 

In order to study the effect of the Fourier transform system 4f on the propagation properties of Airy beams with initial launch angle we investigate the dynamical propagation for different values of $f_{2}$ where $f_{1}=150mm$ is fixed. The Airy beams is characterize by the following parameters: $\lambda=632.8\text{nm}$, $a=0.1$ and $w_{0}=50\mu m$. We used f2 values of 100mm, 150mm and 200mm to characterize Airy beam parameters when passing through convex lenses of different focal distances and validating the predicted theoretical results.

\begin{figure}[H]
 \centering
 \includegraphics[scale=0.30]{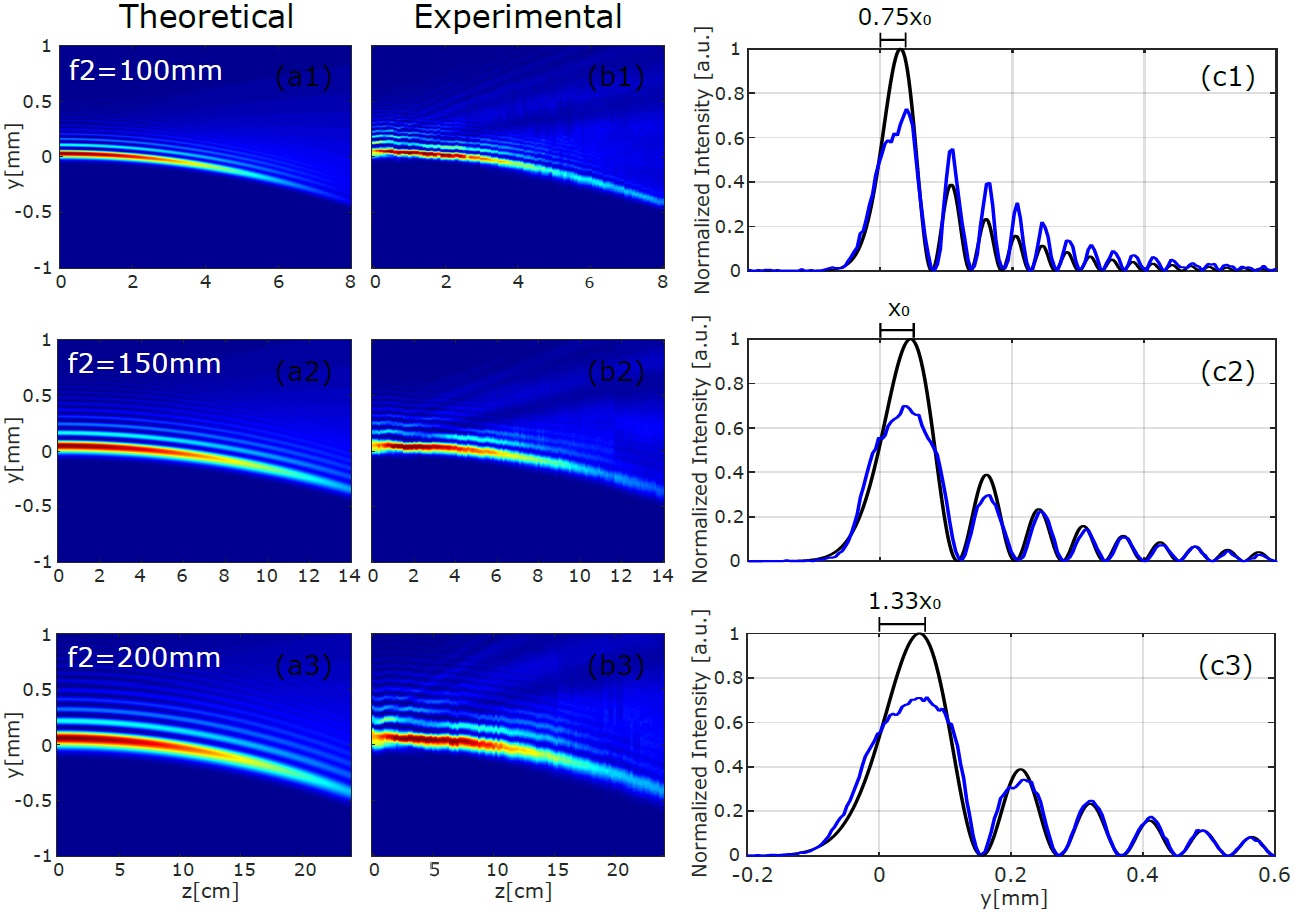}
 \caption{Dynamics propagation theoretical $(a1)$-$(a3)$, obtained using the Eq.~\ref{Airy_ABCD}, and experimental $(b1)$-$(b3)$ as well as, transverse pattern of the intensities on the plane $z=0$ $(c1)$-$(c3)$ for Airy beam one-dimensional with $\nu=0$, at different focal distance $f_{2}=(100,150,200mm)$}
 \label{Dynamics_Propagations_1D}
\end{figure}

\textbf{\textit{First case (Airy beams in $1$D):}} In the Fig.~\ref{Dynamics_Propagations_1D}, the dynamics propagation theoretical $(a1)$-$(a3)$, obtained using the Eq.~\ref{Airy_ABCD}, and experimental $(b1)$-$(b3)$ as well as, transverse pattern of the intensities on the plane $z=0$ $(c1)$-$(c3)$ for Airy beam one-dimensional with $\nu=0$, at different focal distance $f_{2}=(100,150,200mm)$ are shown. It can be easily seen that the propagation through a 4f system conserve the intensity profile, which is governed by an Airy function Eq.~\ref{Airy_ABCD}, where the negative sign before $x$ results in that the lateral side lobes are located at the right side. However, the characteristic length $x_{0}$ becames $\left( f_{2}/f_{1}\right) x_{0}$ as can be seen in the Fig.~\ref{Dynamics_Propagations_1D} $(c1)$-$(c3)$. Also, It could be noted that the diffraction-free propagation and spot of the Airy beams increases with  $f_{2}$. 

\begin{figure}[H]
 \centering
 \includegraphics[scale=0.30]{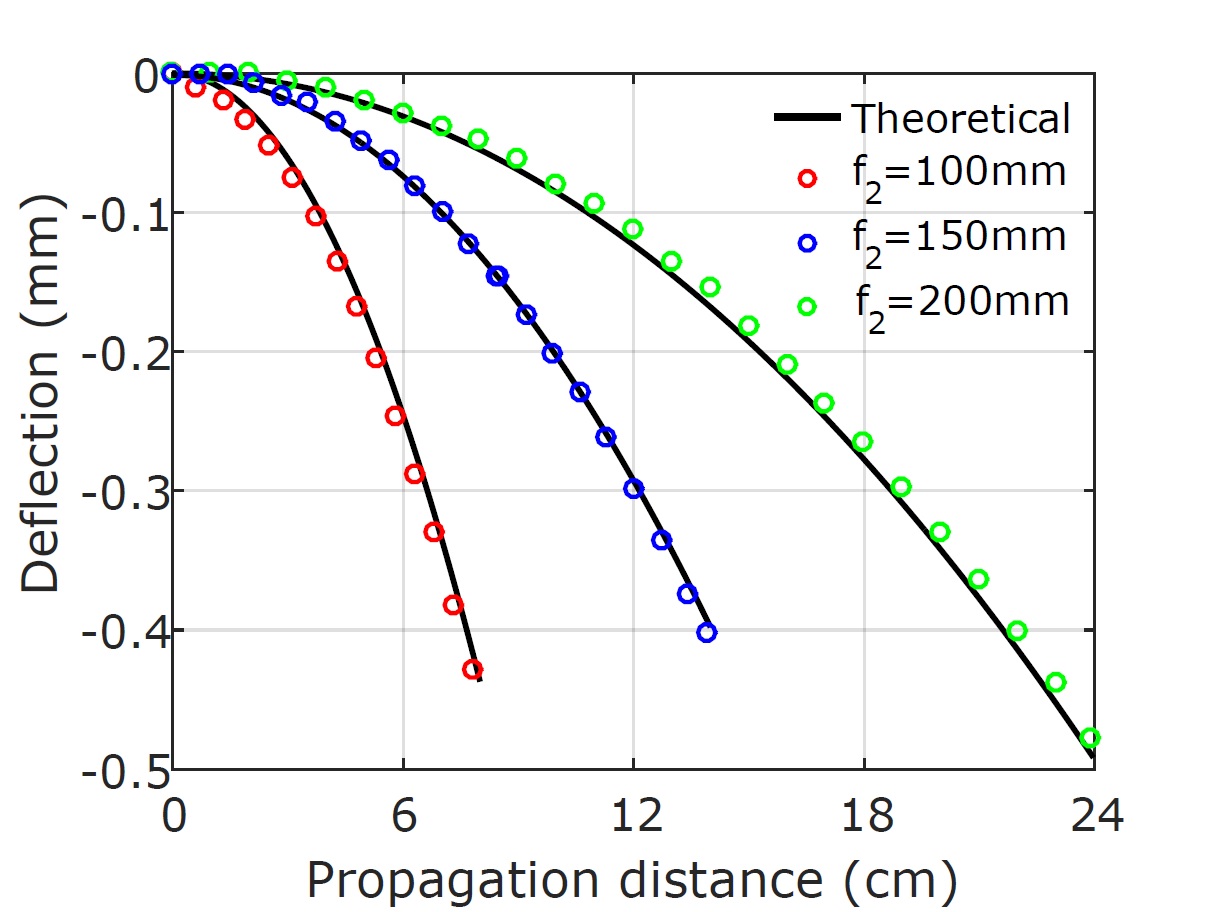}
 \caption{Deflections of the Airy beam as a function of the distance of propagation for $\nu=0$ and different values of $f_{2}$}
 \label{Deflection}
\end{figure}

The Fig.~\ref{Deflection} shows the deflections of the Airy beam as a function of the distance of propagation for different values of $f_{2}$. The parabolic trajectory is described by the theoretical relation Eq.~\ref{Balisctic_Trajectory}. We can see that deflection coefficient, given by $b_{0}=\left(f_{1}/f_{2}\right)^3\left(1/4k^2x_{0}^3\right)$, has a cubic variation with the ratio between the focal distances, resulting in an decrease in the deflection coefficient with $f_{2}$.

\begin{figure}[H]
 \centering
 \includegraphics[scale=0.32]{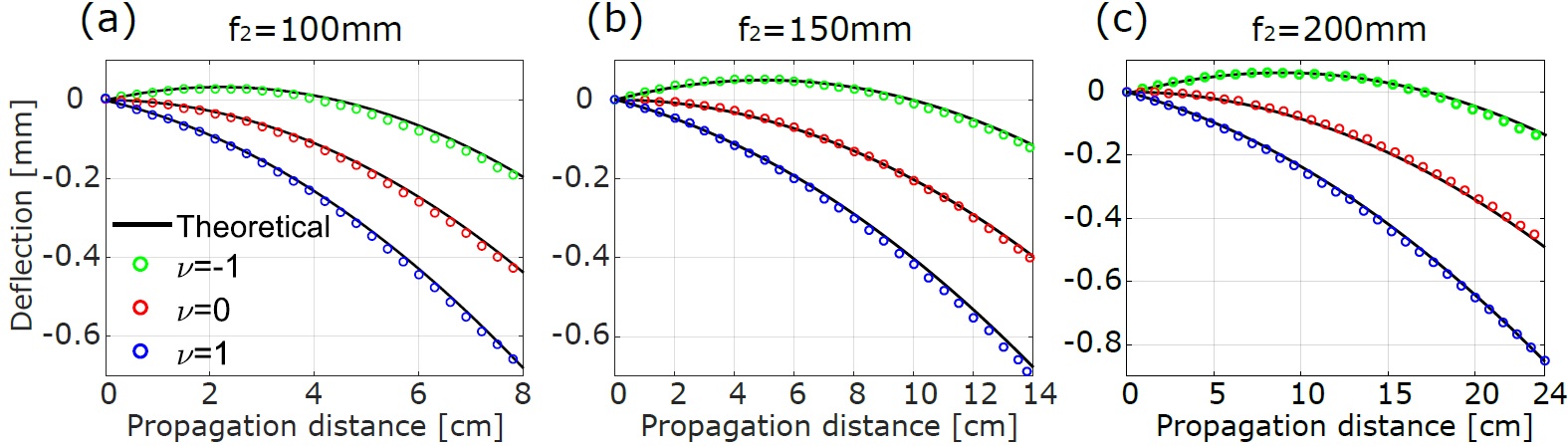}
 \caption{Deflections of the Airy beam as a function of the distance of propagation for different values of $\nu$ in $(a)$ $f_{2}=100mm$, $(b)$ $f_{2}=150mm$, and $(c)$ $f_{2}=200mm$.}
 \label{Deflection_Different_f}
\end{figure}

In the Fig~\ref{Deflection_Different_f} we can see the deflection for different values of $\nu$. From the Eq.~\ref{Balisctic_Trajectory} one can see that that the initial launch angles as well as the focal distance $f_{2}$ can be used to obtain a optimal control of the ballistic trajectory. For $\nu<0$, We have that the maximum point of deflection happens when $x_{max}=\left(f_{2}/f_{1}\right)x_{0}\nu^2$, thus having a linear behavior with $f_{2}$ and quadratic with launch angles $\nu$. We can see that the results obtained experimentally are in agreement with those predicted theoretically.
\textbf{\textit{Second case (Airy beams in $2$D):}} The results for $1$D can be generalized in $2$D taking the scalar field what described the beam as the product of two independent components \cite{Sivilo2007,Siviloglou2008}, that is:
\begin{equation}
\psi\left(s_{x},s_{y},\xi_{x},\xi_{y}\right)=\psi_{x}\left(s_{x},\xi_{x}\right)\psi_{y}\left(s_{x},\xi_{y}\right)\,,
\label{Airy_2D}
\end{equation}
where each of the components $\psi_{x}\left(s_{x},\xi_{x}\right)$ e $\psi_{y}\left(s_{x},\xi_{y}\right)$ satisfies the Eq.~\ref{paraxial_equation} and is given by Eq.~\ref{Airy_2D}, with $s_{x}=x/x_{0}$, $s_{y}=y/y_{0}$, $\xi_{x}=z/kx_{0}^{2}$ e $\xi_{y}=z/ky_{0}^{2}$. To simplify the Airy beams description, we will consider a symmetrical configuration, such as, $a_{x}=a_{x}=a$, $\nu_{x}=\nu_{y}=\nu$ and $x_{0}=y_{0}=w_{0}$, resulting in $\xi_{x}=\xi_{y}=\xi=z/kw_{0}^{2}$.

\begin{figure}[H]
 \centering
 \includegraphics[scale=0.50]{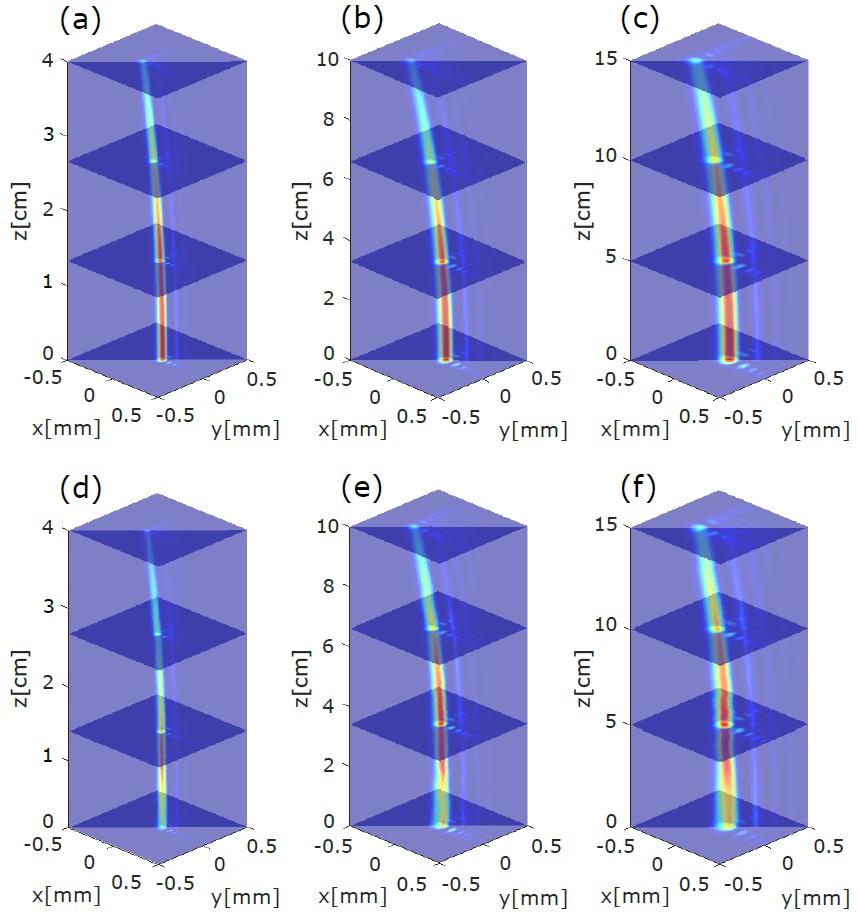}
 \caption{Dynamic propagation of the intensity distribution theoretical (top row) and experimental (bottom row) for different values of $f_{2}$. $f_{2}=100mm$ in $(a)$ e $(d)$, $f_{2}=150mm$ in $(b)$ e $(e)$, and $f_{2}=200mm$ in $(c)$ e $(f)$.   }
 \label{Propagation_Airy}
\end{figure}

In Fig.~\ref{Propagation_Airy} we can see the dynamic propagation of the normalized intensity distribution of Airy beams two-dimensional, theoretical (top row, Fig.~\ref{Propagation_Airy} $(a)$-$(c)$) and experimental (bottom row, Fig.~\ref{Propagation_Airy} $(d)$-$(f)$), when $\nu=0$ and different values of $f_{2}$. The Intensity cross section for different propagation distance, exhibit the dynamic transverse profile. In this case, the beam conserved the same properties in the propagation of the beam one-dimensional, and this  is accelerated  along of $45^{0}$ axis in the $x$-$y$ plane where the parabolic trajectory is described by $x_{d}=y_{d}=(f_{1}/f_{2})^3(z^2/4k^2 w_{0}^3)$.

\begin{figure}[H]
 \centering
 \includegraphics[scale=0.36]{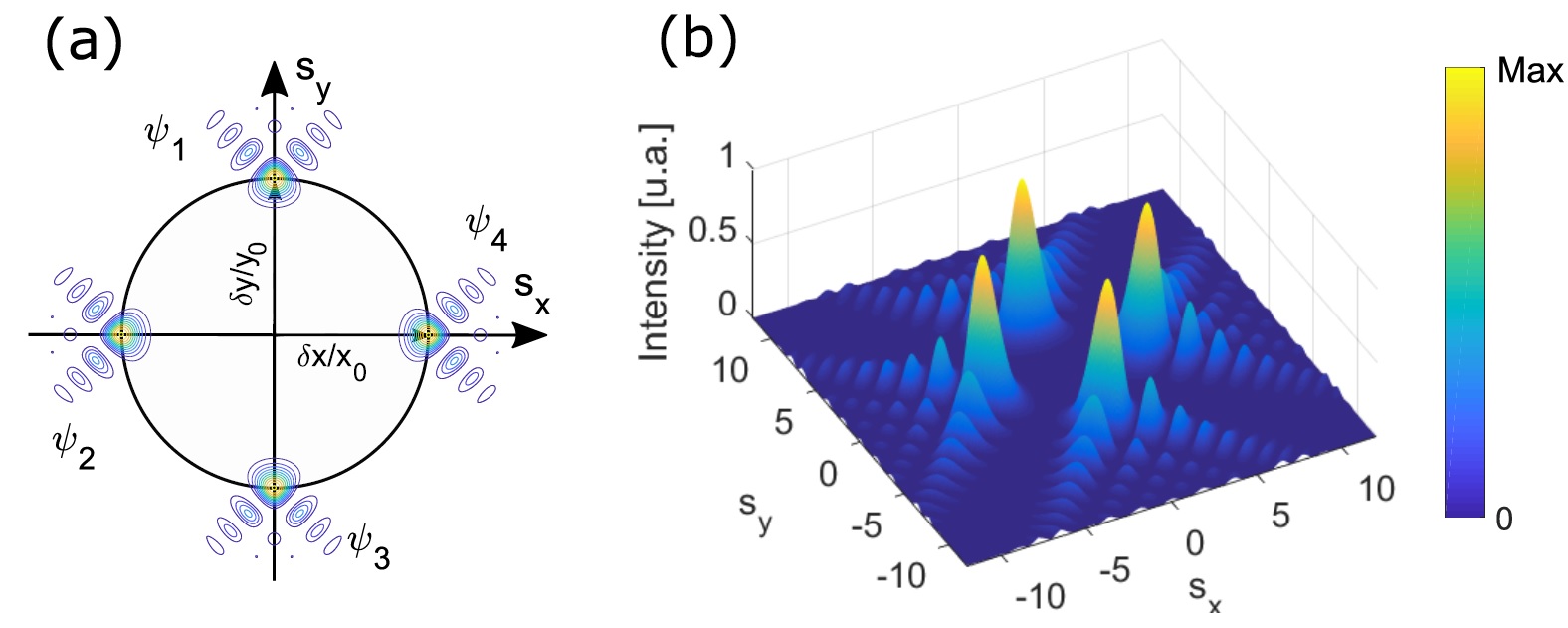}
 \caption{$(a)$ Scheme for the generation of AAiBs with four symmetrical Airy beam ($N=4$). $(b)$ Transverse intensity pattern from the plane $\xi=0$ onwards.}
 \label{scheme}
\end{figure}

\textbf{\textit{Third case (Array of Airy beams):}} We consider $N$ even Airy beams spatially displaced on the transverse plane in $(\delta x,\delta y)$, an array of Airy beams (AAiBs). Through the $\theta=2\pi/N + 3\pi/4$ degree rotation of Eq.~\ref{Airy_2D} in the transverse plane, we can get the set of $N$ rotated symmetrical Airy beams which accelerate mutually in the opposite direction \cite{Cheng2014,Lu2017,Suarez2016}. The total field is given by, 
\begin{equation}
\Psi\left(x,y,z\right) =\sum_{j=1}^{N}\psi_{jx}\left(s_{jx},s_{jy},\xi\right)\psi_{jy}\left(s_{jx},s_{jy},\xi\right)\,,
\label{matriz_superpositions}
\end{equation}
where 
\begin{equation}
\begin{split}
x_{j}&=x\text{cos}\theta_{j}-y\text{sin}\theta_{j}-\dfrac{f_{2}}{f_{1}}\delta x\\
y_{j}&=x\text{sin}\theta_{j}+y\text{cos}\theta_{j}-\dfrac{f_{2}}{f_{1}}\delta y\,,
\end{split}
\label{matriz_superpositions1}
\end{equation}
The angle $\theta_{j}=2(j-1)\pi/N + 3\pi/4$ denotes the angle of rotation around the $z$ axis. In the Fig.~\ref{scheme} we can see a representative scheme of the generation process of an array of Airy beams (AAiBs) from four symmetrical Airy beam propagating along the z axis \cite{Suarez2016}.

\begin{figure}[H]
 \centering
 \includegraphics[scale=0.50]{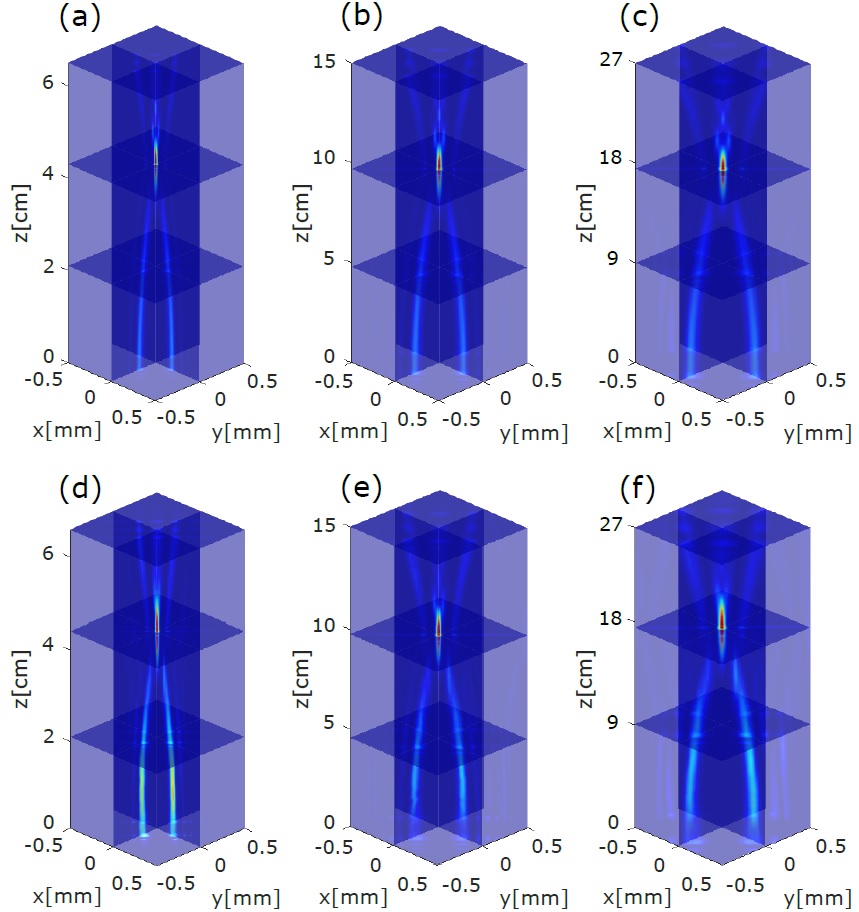}
 \caption{Dynamic propagation of the intensity distribution theoretical (top row) and experimental (bottom row) for different values of $f_{2}$. $f_{2}=100mm$ in $(a)$ e $(d)$, $f_{2}=150mm$ in $(b)$ e $(e)$, and $f_{2}=200mm$ in $(c)$ e $(f)$.  }
 \label{Propagation_Matrix_Airy_3D}
\end{figure}

The Fig.~\ref {Propagation_Matrix_Airy_3D} shows the dynamic propagation of the normalized intensity distribution of a AAiBs  theoretical (top row, Fig.~\ref{Propagation_Matrix_Airy_3D} $(a)$-$(c)$) and experimental (bottom row, Fig.~\ref{Propagation_Matrix_Airy_3D} $(d)$-$(f)$) when $\nu=0$ and different values of $f_{2}$. The AAiBs is formed by $N=4$ Airy beams where each beam is characterized by the same parameter with $\delta x=\delta y=150\mu m$. The intensity cross section for different propagation distance, exhibit the dynamic transverse profile.     

We can see, that the $4$ Airy beam in the array are propagated along of the $z$ axis and converge symmetrically at the same point due to parabolic trajectory which is described by Eq.~\ref{Balisctic_Trajectory}. We can observed as the focal point, $z_{f}=-2kw_{0}(f_{2}/f_{1})^{2}\left(\nu - \sqrt{\nu^2+s_{0}/w_{0}} \right)$, which is defined by is defined by the common point where the main lobes from all the beams in the array intercept along of the beam, where $s_{0}$ is the position of the first intensity peak, have a quadratic variation whit the ratio between the focal distance, resulting in a increase of the focal point whit $f_{2}$. 

\begin{figure}[H]
 \centering
 \includegraphics[scale=0.30]{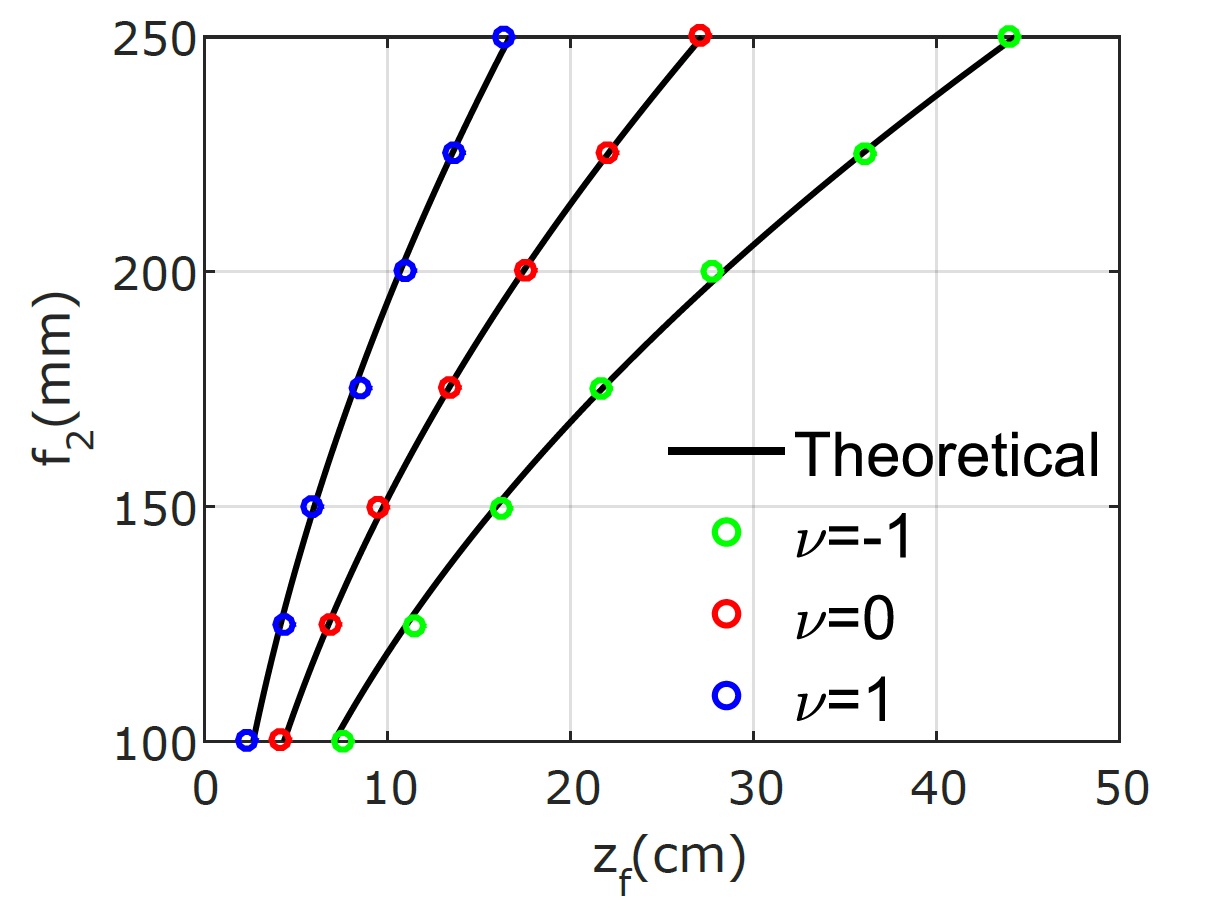}
 \caption{Intensity profile along the axis of propagation theoretical (black line) and experimental (blue point) under different values of $f_{2}$. $(a)$ $f_{2}=100mm$, $(b)$ $f_{2}=150mm$, and $(c)$ $f_{2}=200mm$.}
 \label{Focal_ponit}
\end{figure}

The Fig~\ref{Focal_ponit} shows the behavior of the focal point as a function of $f_{2}$ for three different values of the $\nu$. These results imply that we have an control of the auto-focusing properties of the Airy beams simply by varying the initial angle launch or the focal distance of $f_{2}$ or both parameters. 
 
\h {\em 4. Conclusions} --- We validate theoretical and experimental study of the propagation of Airy beams when crossing through different lenses in a holographic reconstruction system of computer generated holograms (CGH) implemented electronically in a spatial light modulator (SLM). We analytically and experimentally showed that, when a Airy beam propagates through a two convex lens system, the beam that emanates from the second lens is again a Airy beam, which however suffered important alterations: its transverse intensity pattern size is scaled by the lens focal ratio, $(f_{1}/f_{2})$; and, the deflection coefficient has a cubic variation with the ratio between the focal distances resulting in an  change in the deflection coefficient scaled by $(f_{1}/f_{2})^3$. The implementation of amplitude CGHs over LC-SLM for generation of Airy beam in 1D, 2D and array of Airy beams (AAiBs) with finite energy crossing different lenses allows controlling the transverse and longitudinal dimensions of the beams as well as its deflection. The results here presented are in agreement with the theoretical model and can be useful in the generation of structured Airy beams within very small spatial regions. Also its open exciting possibilities of generate and control potentially interesting Airy beams for scientific and technological applications: optical tweezers, plasma physics, optical microscopy, optical communications and optical metrology. \\

\h {\em Acknowledgments} The authors acknowledge partial support from UFABC, CAPES, FAPESP (UNDER GRANTS 16/19131-6) and CNPq (UNDER GRANTS 302070/2017-6).

\end{document}